\documentclass[a4paper,fleqn,usenatbib]{mnras}

\usepackage[T1]{fontenc}
\usepackage{ae,aecompl}

\usepackage{graphicx}	
\usepackage{amsmath,mathtools}	
\usepackage{amssymb}	

\newcommand{\cmt}{cm$^{-3}$}

\newcommand{\s}{s$^{-1}$}

\title[The Nautilus 3-phase model]{Gas and grain chemical composition in cold cores as predicted by the Nautilus 3-phase model}

\author[M. Ruaud et al.]{
Maxime Ruaud,$^{1,2}$\thanks{E-mail: ruaud@obs.u-bordeaux1.fr}
Valentine Wakelam,$^{1,2}$
and Franck Hersant$^{1,2}$
\\
$^{1}$Univ. Bordeaux, LAB, UMR 5804, F-33270, Floirac, France\\
$^{2}$CNRS, LAB, UMR 5804, F-33270 Floirac, France
}

\date{Accepted 2016 April 13. Received 2016 April 07; in original form 2016 January 26}

\pubyear{2015}

\begin{document}
\label{firstpage}
\pagerange{\pageref{firstpage}--\pageref{lastpage}}
\maketitle

\begin{abstract}
We present an extended version of the 2-phase gas-grain code {\sevensize NAUTILUS} to the 3-phase modelling of gas and grain chemistry of cold cores. In this model, both the mantle and the surface are considered as chemically active. We also take into account the competition among reaction, diffusion and evaporation. The model predictions are confronted to ice observations in the envelope of low-mass and massive young stellar objects as well as toward background stars. Modelled gas-phase abundances are compared to species observed toward TMC-1 (CP) and L134N dark clouds. We find that our model successfully reproduces the observed ice species. It is found that the reaction-diffusion competition strongly enhances reactions with barriers and more specifically reactions with H$_2$, which is abundant on grains. This finding highlights the importance to have a good approach to determine the abundance of H$_2$ on grains. Consequently, it is found that the major N-bearing species on grains go from NH$_3$ to N$_2$ and HCN when the reaction-diffusion competition is accounted. In the gas-phase and before few $10^5$ yrs, we find that the 3-phase model does not have a strong impact on the observed species compared to the 2-phase model. After this time, the computed abundances dramatically decrease due to the strong accretion on dust, which is not counterbalanced by the desorption less efficient than in the 2-phase model. This strongly constrains the chemical-age of cold cores to be of the order of few $10^5$ yrs.
\end{abstract}


%
\section{Introduction}
Interstellar grains play a central role in the physics of the interstellar medium. These grains are responsible for the observed extinction of the starlight \citep{Trumpler30} and provide shielding, allowing an increase of the molecular complexity in dense regions. They are also known to act as catalysts for reactions, the most important one leading to the formation of H$_2$ \citep{Gould63}. In dense and shielded regions, icy mantles are known to form around dust grain cores via accretion of gas-phase species followed by chemical reactions at the surface of the grains. Observations of these mantles have shown that they mostly consist of amorphous water, carbon monoxide, carbon dioxide, methanol, ammonia and methane \citep[see the review of][]{Boogert15}. Although difficult to identify, several other compounds are thought to be present and formed on the surface of those grains. In particular, the observed gas-phase abundance of complex molecules in star-forming environments seems tightly linked to the physical and chemical processes occurring in the solid state \citep[see the review of][]{Herbst09}. However, the recent detection of molecules such as methyl formate or dimethyl ether in cold regions \citep{Cernicharo12,Bacmann12} have challenged models and brought some authors \citep[see][for example]{Balucani15} to reassess the relative contribution of gas phase reactions compared to grain chemistry. Although the relative contribution of gas phase and grain surface chemistry on the formation of the most complex organic molecules is still unclear, it seems well established that grains plays a crucial role even at low temperature. This is the case for the formation of methanol for example for which gas phase chemistry cannot alone explain its observed abundances \citep{Geppert06,Garrod06a}. Modelling the formation and the evolution of these icy grains is therefore essential if we aim at studying the chemistry of dense and cold regions.
 
A number of studies has addressed this link by coupling the gas-phase chemistry to the solid phase using various numerical methods with different degrees of complexity and refinement. The four usually employed approaches are: (1) the rate equation approach \citep{Hendecourt85,Hasegawa92,Caselli98}, (2) the master equation \citep{Biham01,Stantcheva02}, (3) the macroscopic kinetic Monte Carlo method \citep{Tielens82,Vasyunin13}, and (4) the microscopic kinetic Monte Carlo method \citep{Chang05,Cuppen05}. Rate equation methods are relatively easy to implement and fast in term of computation time \citep[see][for a review on such techniques]{Cuppen13}.

The model developed by \citet{Hasegawa92} refers to the well known 2-phase (gas and grains) model in which no distinction is made between the outermost and inner monolayers. In this model, the surface and the mantle are treated as a single phase with the same properties for adsorption, diffusion and reaction. Subsequently, \citet{Hasegawa93a} developed a 3-phase (gas, grain surface and grain mantle) model in which the surface and the mantle are treated as separate phases. In this model, species can only react and desorb from the surface phase. The mantle is considered as inert and grows by incorporating material from the surface. Following the same approach, \citet{Taquet12} developed a similar model but using a multilayer method in which each layer is considered as a specific entity rather than one single, averaged, phase, i.e. the mantle. Based on recent calculations and experiments it has become clear that sub-surface processes, such as photodissociation, diffusion and recombination should be at work in interstellar grain mantles \citep{Andersson08,Oberg09c}. Motivated by these findings several authors modified the original 3-phase model of \citet{Hasegawa93a} to take into account this active bulk chemistry \citep{Kalvans10,Garrod13}.

In this paper we present an extended version of the 2-phase gas-grain code {\sevensize NAUTILUS} to the 3-phase modelling. The formalism used to distinguish surface and mantle layers is based on the pioneering work of \citet{Hasegawa93a}. In this model, both the mantle and the surface are considered as chemically active. As opposed to previous models, we assume that the bulk diffusion is driven by the diffusion of water molecules in the ice, as suggested by molecular dynamics simulations and experiments \citep{Ghesquiere15}. This paper is organized as follows. In Section \ref{sec:model_description}, we describe our chemical model and all the processes that we take into account. In Section \ref{sec:ice_modelisation}, we present our modelling results for the solid phase under dark cloud conditions and we compare these results with the available observations. In Section \ref{sec:gas_modelisation}, we present our modelling results for the gas-phase and we compare these results with observations in the TMC-1 (CP) and L134N dark clouds. Finally, Section \ref{sec:conclusion} contains a summary of our work.

%
\section{Model description}
\label{sec:model_description}
This new 3-phase model (gas, grain surface and grain mantle) is a modified version of the 2-phase gas-grain code {\sevensize NAUTILUS} extensively used to study the chemistry in various environments \citep{Loison14,Wakelam14,Reboussin15} and already presented in details in \citet{Reboussin14} and \citet{Ruaud15}. As in \citet{Hasegawa93a}, it is assumed that both accretion and desorption only occur between the gas and the outermost surface layer. We have however assumed that chemical reactions can occur in the mantle as on the surface but with a smaller diffusion rate of the species compared to the surface.

\subsection{General equations}
\label{sec:general_eq}
The differential equations governing gaseous, surface and mantle concentrations of a species i, denoted as  $n(i)$, $n_s(i)$ and $n_m(i)$ [\cmt] respectively, are:

\begin{equation*}
\begin{split}
\frac{dn(i)}{dt}\biggr\rvert _\text{tot} &=\sum_l \sum_j k_{lj}n(l)n(j) + k_\text{diss}(j)n(j) + k_\text{des}(i)n_s(i) \\ 
& - k_\text{acc}(i)n(i) - k_\text{diss}(i)n(i) - n(i)\sum_l k_{ij} n(j),
\end{split}
\end{equation*}

\begin{equation*}
\begin{split}
\frac{dn_s(i)}{dt}\biggr\rvert _\text{tot}&=\sum_l \sum_j k_{lj}^sn_s(l)n_s(j) + k_\text{diss}^s(j)n_s(j) + k_\text{acc}(i)n(i)\\
& + k_\text{swap}^{m}(i) n_m(i)  + \frac{dn_m(i)}{dt} \biggr\rvert _{m\rightarrow s} \\
& - n_s(i)\sum_j k_{ij}^s n_s(j) -  k_\text{des}(i)n_s(i) - k_\text{diss}^s(i)n_s(i) \\
& - k_\text{swap}^{s}(i) n_s(i) - \frac{dn_s(i)}{dt} \biggr\rvert _{s\rightarrow m},
\end{split}
\end{equation*}

\begin{equation*}
\begin{split}
\frac{dn_m(i)}{dt}\biggr\rvert _\text{tot}&=\sum_l \sum_j k_{lj}^m n_m(l)n_m(j) + k_\text{diss}^m(j)n_m(j) \\
& + k_\text{swap}^{m}(i) n_s(i)  + \frac{dn_s(i)}{dt} \biggr\rvert _{s\rightarrow m}\\
& - n_m(i)\sum_j k_{ij}^m n_m(j) - k_\text{diss}^m(i)n_m(i) \\
&- k_\text{swap}^{m}(i) n_m(i) - \frac{dn_m(i)}{dt} \biggr\rvert _{m\rightarrow s}.
\end{split}
\end{equation*}
$k_{ij}$, $k_{ij}^s$ and $k_{ij}^m$ [cm$^3$\s] represent the rate coefficients for reactions between species $i$ and $j$ in the gas-phase, on the grain surface and in the mantle respectively. $k_\text{acc}$ and $k_\text{des}$ [\s] are respectively the accretion/evaporation rate coefficient of individual species onto/from the grain surface. In our model $k_\text{des}$ can be thermal or non-thermal (see section \ref{sec:acc_diff_des}). $k_\text{diss}$ [\s] gather dissociation rate coefficients by (1) direct  cosmic-ray processes, (2) secondary UV photons induced by cosmic-rays and (3) the standard interstellar UV photons \citep[see][]{Wakelam12}. $k_\text{swap}^s$ and $k_\text{swap}^m$ [\s] are the surface-mantle and mantle-surface swapping rates respectively. $dn_s(i)/dt\rvert_{s\rightarrow m}$ and $dn_m(i)/dt\rvert_{m\rightarrow s}$ are related to the individual transfer of the species $i$ from the surface to the mantle and \textit{vice versa}. Those terms are calculated as follow: 

\begin{equation*}
\frac{dn_s(i)}{dt} \biggr\rvert _{s\rightarrow m} = \alpha_\text{gain} \frac{n_s(i)}{n_{s,\text{tot}}} \frac{dn_{s,\text{gain}}}{dt},
\end{equation*}
and
\begin{equation*}
\frac{dn_m(i)}{dt} \biggr\rvert _{m\rightarrow s} = \alpha_\text{loss} \frac{n_m(i)}{n_{m,\text{tot}}} \frac{dn_{s,\text{loss}}}{dt},
\end{equation*}

where $dn_{s,\text{gain}}/dt$ and $dn_{s,\text{loss}}/dt$ are the overall rate of gain and loss of surface material respectively,
\begin{equation*}
\begin{split}
\frac{dn_{s,\text{gain}}}{dt} &= \sum_i \Bigg[ \sum_l \sum_j k_{lj}^sn_s(l)n_s(j) + k_\text{diss}^s(j)n_s(j)\\ 
& + k_\text{acc}(i)n(i) \Bigg],
\end{split}
\end{equation*}

\begin{equation*}
\begin{split}
\frac{dn_{s,\text{loss}}}{dt} &=  \sum_i \Bigg[ n_s(i)\sum_j k_{ij}^s n_s(j) +  k_\text{des}(i)n_s(i) \\ 
& + k_\text{diss}^s(i)n_s(i)\Bigg].
\end{split}
\end{equation*}
Therefore, the sum of $dn_{s,\text{gain}}/dt$ and $dn_{s,\text{loss}}/dt$ represent the net rate of change in the total surface material. This sum is thus $>0$ in the case of net accretion from the gas phase onto grain surface and $<0$ in the case of net desorption from the grain surface to the gas phase. This approach slightly differs to the one proposed by \citet{Hasegawa93a} in the sense that the overall rate of gain and loss of surface molecules are simultaneously used to compute the individual rate of species transfer from the surface to the mantle and from the mantle to the surface respectively. Using this notation, we can define the average absolute number of particles of species $i$ present in the surface layer as $N_s(i)= n_s(i) / n_{\text{dust}}$ where $n_s(i)$ [\cmt] is the surface concentration of surface species $i$ and $n_{\text{dust}}$ [\cmt] is the gas-phase concentration of dust particles.
$\alpha_\text{gain}$ and $\alpha_\text{loss}$ are defined as \citep[see][]{Garrod11}
\begin{equation*}
\alpha_\text{gain}= \frac{\sum_i N_{s}(i)}{\beta N_\text{site}}
\end{equation*}
and
\begin{equation*}
\alpha_\text{loss}=
\begin{dcases}
\frac{\sum_i N_m(i)}{\sum_i N_s(i)},  &\text{if}~~\sum\nolimits_i N_{m}(i) < \sum\nolimits_i N_{s}(i)\\
1,& \text{otherwise}
\end{dcases}
\end{equation*}
where $N_\text{site}$ is the total number of surface sites on a grain. $\beta$ is used to set the number of active surface layers on the grain. We assume that $\beta=2$ as suggested by \citet{Fayolle11} to account for surface roughness; i.e. the two outermost layers are considered as part of the surface.
It is important to note that since $dn_s(i)/dt\rvert_{s\rightarrow m}$ and $dn_m(i)/dt\rvert_{m\rightarrow s}$ are related to the net rate of change in total surface material, those terms do not represent a physical motion of material between surface and mantle but rather a way to treat the continuous renewal of the grain surface material \citep{Hasegawa93a,Garrod11}. The true physical motion of material between surface and mantle is accounted by 
$k_\text{swap}^{s}$ and $k_\text{swap}^{m}$ [\s]. As discussed by \citet{Garrod13}, the surface-mantle swapping is a pair-wise process and requires that each swapping rate of all species, must collectively produce no net transfer between the two phases. In our model this requirement is achieved following \citet{Garrod13}. We first calculate the swapping rates from mantle to surface,

\begin{equation*}
k_\text{swap}^{m} (i)=
\begin{dcases}
\frac{1}{t_\text{hop}^m(i)} [\text{s}^{-1}] & \text{if} ~~ N_{\text{lay},m} < 1\\
\frac{1}{t_\text{hop}^m(i) N_{\text{lay},m}} [\text{s}^{-1}] & \text{if} ~~ N_{\text{lay},m} > 1\\
\end{dcases}
\end{equation*}
where $N_{\text{lay},m} = \sum_i N_m(i)/N_\text{site}$ is the number of mantle layers and $t_\text{hop}^m = \nu^{-1} \exp({E_\text{diff}^m/T_\text{dust}})$ the mantle thermal hopping time (i.e. the time required for the species to migrate from one site to an adjacent one via thermal hopping).
The surface to mantle rates are then built to match a total rate identical to the mantle to surface one,
\begin{equation*}
k_\text{swap}^{s}(i)=\frac{\sum_j k_\text{swap}^{m}(j)n_m(j)}{n_{s,\text{tot}}} [\text{s}^{-1}],
\end{equation*}

Unlike \citet{Garrod13}, we set all the species (except H, H$_2$, C, N and O) with E$_\text{diff}^m(i)$$<$E$_\text{diff}^m$(H$_2$O) to the diffusion barrier of H$_2$O considering that the bulk diffusion mechanism is driven by the diffusion of the water molecules in the ice, in agreement with the calculations of \citet{Ghesquiere15} for NH$_3$, CO$_2$, CO and H$_2$CO trapped in an amorphous H$_2$O ice for temperatures in the range of 90-170K. This treatment allows us to reproduce the observed desorption kinetics by TPDs experiments for which first and zero order desorption kinetics are at work (See Appendix \ref{sec:tpd_expe}). We do not include H, H$_2$, C, N and O assuming that these species are sufficiently light to efficiently diffuse in the ice even at low temperature.

\subsection{Accretion, diffusion and desorption}
\label{sec:acc_diff_des}
Accretion, diffusion and thermal desorption rate coefficients are computed based on \citet{Hasegawa92}. We assume a sticking probability of 1.0 for all the neutral species, except for H and H$_2$. For these two species, we use the temperature dependent expressions of \citet{Matar10} and \citet{Chaabouni12} derived from laboratory experiments. As stated in section \ref{sec:general_eq}, several non-thermal desorption mechanisms are also included: (1) the cosmic-rays desorption mechanism \citep[see][]{Hasegawa93b}, (2) the chemical desorption mechanism (desorption due to the exothermicity of surface reactions, set to $\sim 1\%$ efficiency) \citep[see][]{Garrod07}, and (3) the photo-desorption. The non thermal desorption mechanisms considered in other studies such as \citet{Roberts07} are not included in this work.

Over the last years, quantitative photo-desorption yields have been measured experimentally for pure ices sample using hydrogen discharge lamps: CO \citep{Oberg07,MunozCaro10}, H$_2$O \citep{Oberg09b}, N$_2$, CO$_2$ \citep{Oberg09a,Yuan13} and CH$_3$OH \citep{Oberg09c}. Thanks to these experiments, useful photo-desorption yields were determined for astrochemical models despite the fact that the mechanism itself was not very well understood. More recently, wavelength-dependent UV studies have been performed for pure and mixed ices: CO \citep{Fayolle11a,Bertin12,Bertin13}, N$_2$ and O$_2$ \citep{Fayolle13}. These wavelength-dependent studies helped in the understanding of the photo-desorption process and give new insights on this process. From these experiments, it seems to emerge that:
\begin{itemize}
\item The photo-desorption is an indirect process: induced by a photo absorption in sub-surface molecular layers and an energy transfer to the surface molecules \citep{MunozCaro10,Bertin12}. 
\item Only the topmost ice layers are involved in the photo-desorption process; i.e. the 2 to 5 first monolayers.
\item The photo-desorption process is highly wavelength dependent as shown in \citet{Fayolle11a,Bertin12,Bertin13} and \citet{Fayolle13}.
\item This mechanism highly depends on the composition of the ice and how this ice is mixed, as shown by \citet{Bertin12} and \citet{Bertin13}. 
\end{itemize}

Following this and considering the lack of knowledge on this process, we follow the recommendations of \citet{Bertin13} and choose to use a simplistic approach and to consider a single photo-desorption yield for all the molecules rather than individual ones determined from experiments. In the following the photo-desorption rate is kept constant using a photo-desorption yield $Y_\text{pd} = 10^{-4}$ molecule/photon \citep{Andersson08}. The corresponding rates for (1) the standard interstellar UV photons and (2) the secondary UV photons induced by cosmic-rays used are
\begin{equation*}
k_\text{des,UV}(i)= F_\text{UV}S_\text{UV}  \exp({-2 A_V})Y_\text{pd}\frac{\pi r_\text{dust}^2}{N_\text{site}}~~[\text{s}^{-1}],
\end{equation*}
for the standard interstellar UV photons and 
\begin{equation*}
k_\text{des,UV-CR}(i)= F_\text{UV-CR} S_\text{UV-CR}Y_\text{pd}\frac{\pi r_\text{dust}^2}{N_\text{site}}~~[\text{s}^{-1}],
\end{equation*}
for the secondary UV photons induced by cosmic-rays. In those expressions, $S$ is the scaling factor for the UV radiation field, $ F $ its strength in unit of photons cm$^{-2}$ \s and $r_\text{dust}$ the grain radius. The factor of 2 in the exponential part of the first expression comes from \citet{Roberge91} and account for the UV extinction relative to $A_V$. For the standard interstellar UV photons field, we set $F_\text{UV}=1.0\times 10^8$ photons cm$^{-2}$ \s~\citep{Oberg07} and $F_\text{UV-CR}=1.0\times 10^4$ photons cm$^{-2}$ \s~\citep{Shen04}. In the following, the impact of the photo-desorption is not discussed since it does not impact our modelling results.

\subsection{Grain surface reactions}
We consider that most reactions proceed through the Langmuir-Hinshelwood mechanism. In a previous work, we also considered that some reactions can proceed through the Eley-Rideal mechanism \citep[see][for details]{Ruaud15} but did not consider this mechanism here. The reason is that this mechanism is found to have a minor impact on the computed abundances due to the low diffusion to binding ratio that we use (i.e. $E_\text{diff}^s=0.4\times E_\text{des}$ for our standard model, see Section \ref{sec:ice_modelisation}). We carried tests to see the impact of the  Eley-Rideal mechanism and found that for $E_\text{diff}^s \gtrsim 0.45\times E_\text{des}$ this mechanism compete with the Langmuir-Hinshelwood mechanism due to the higher surface diffusion energy. When $E_\text{diff}^s \gtrsim 0.45\times E_\text{des}$ this mechanism enhances the computed abundances of species like CH$_3$OCH$_3$, HCOOCH$_3$ and CH$_3$CHO \citep[as already found in][]{Ruaud15}.

The Langmuir-Hinshelwood mechanism supposes that each reactant is already pre adsorbed on the surface (i.e. physisorbed) before encountering another reactant and possibly leading to the formation of a new species. The surface reaction rate for this mechanism is

\begin{equation*}
k_{ij}^x = \kappa_{ij}\Bigg(\frac{1}{t_\text{hop}^x(i)} + \frac{1}{t_\text{hop}^x(j)}\Bigg)\frac{1}{N_\text{site}n_\text{dust}}~ [\text{cm}^3\text{s}^{-1}],
\end{equation*}
where $x$ stands for $s$ (surface species) and $m$ (mantle species). $\kappa_{ij}$ is the probability that the reaction occurs and $t_\text{hop}^x = \nu_i^{-1} \exp({E_\text{diff}^x(i)/T_\text{dust}})$ the thermal hopping time. In this expression, $\nu_i$ is the characteristic vibration frequency of the species $i$ \citep[see][]{Hasegawa92} and $E_\text{diff}^x(i)$ its diffusion energy barrier. For mantle reactions, we modified the original expression given by \citet{Hasegawa92} by dividing $k_{ij}^m$ by $\sum_i N_m(i)/N_\text{site}$; i.e. the number of mantle layers, considering that the time needed for a mantle species to scan the entire grain mantle increases as the mantle is growing.

When the reaction is exothermic and barrierless, we consider that $\kappa_{ij}=1$ \citep{Hasegawa92}. Inversely, for exothermic reactions with activation barriers, denoted $E_{\text{A},ij}$, we calculate $\kappa_{ij}$ as the result of the competition among reaction, hopping and evaporation as suggested by \citet{Chang07} and \citet{Garrod11}:
\begin{equation*}
\kappa_{ij} = \frac{\nu_{ij} \kappa_{ij}^*}{\nu_{ij} \kappa_{ij}^* + k_\text{hop}^x(i) + k_\text{evap}(i) + k_\text{hop}^x(j) + k_\text{evap}(j)}.
\end{equation*}
where $\kappa_{ij}^*$ can be expressed as $\exp(-E_{\text{A},ij}/\text{T}_\text{dust})$ or the quantum mechanical probability for tunneling through a rectangular barrier of thickness $a$: $\kappa_{ij}^*=\exp[-2(a/ \hbar)(2\mu E_{\text{A},ij})^{1/2}]$, where $\mu$ is the reduced mass \citep[see][for details]{Hasegawa92}. Here, $\nu_{ij}$ is taken to be equal to the larger value of the characteristic frequencies of the two reactants $i$ and $j$ \citep{Garrod11}.

%
\section{Model predictions for the ice composition}
\label{sec:ice_modelisation}

Throughout this paper we use typical cold dense cloud conditions, i.e. n$_\text{H}=2\times10^4$\cmt, T$_\text{gas}$=T$_\text{dust}$=10K, A$_\text{V}$=10 mag and $\zeta_{\text{H}_2}=1.3\times10^{-17}$\s. The adopted gas-phase network is \textit{kida.uva.2014} \citep[see][]{Wakelam15}. The grain network is the one presented in \citet{Ruaud15} and based on the previous work of \citet{Garrod07}. Hydrogen-abstraction reactions were added for reactions with HCO, CH$_3$O, CH$_2$OH and CH$_3$OH \citep{Chuang16}\footnote{The final network contains 717 species (489 for the gas phase and 228 for the grain surface) coupled through more than 11500 reactions (7509 for the gas phase and more than 4000 for the grain surface).}. For the H-atom abstraction reaction from CH$_3$OH, we use the reaction barriers used in \citet{Garrod13}, i.e. $E_\text{A}=4380$K for the reaction leading to CH$_2$OH and $E_\text{A}=6640$K for the reaction leading to CH$_3$O).

The binding energy of H is taken equal to $E_\text{des}(\text{H})=650$ K as suggested by theoretical calculations on amorphous solid water ice \citep{AlHalabi07}.
For diffusion barriers, we assume $E_\text{diff}^s=0.4\times E_\text{des}$ and $E_\text{diff}^m=0.8\times E_\text{des}$ for surface and mantle species respectively\footnote{Note that, except for H, H$_2$, C, N and O, we have set $E_\text{diff}^m(i)$=$E_\text{diff}^m$(H$_2$O) for all species that have $E_\text{diff}^m(i)$<$E_\text{diff}^m$(H$_2$O), as discussed in Section \ref{sec:general_eq}.}. A specific value $E_\text{diff}^s(\text{H})=230$ K for H is used coming from the calculations of \citet{AlHalabi07}. A similar diffusion barrier was derived by the experiment of \citet{Matar08} for D atoms (i.e. $E_\text{diff}^s(\text{D})=255 \pm 23$ K) as well as \citet{Hama12} for H atoms (i.e. $E_\text{diff}^s(\text{H})=255 \pm 11$ K) on amorphous water ice surfaces. Finally, initial abundances are summarized in Table \ref{initial_abundances}, grains are considered to be spherical with a 0.1$\mu$m radius, a 3g.\cmt~density, $\simeq10^6$ surface sites\footnote{The effect of grain growth which leads to an increase of the number of grain sites is neglected.} and a dust to gas mass ratio of 0.01. 

In the following, all the 2-phase model are run using the same parameters than those used for the 3-phase model, except for the diffusion-to-binding energy ratio which is taken to be equal to $E_\text{diff}=0.5\times E_\text{des}$ (to be consistent with previous studies based on 2-phase modelling). For H, the specific values defined above are employed.

\begin{table}
   \caption{Initial abundances. $^\dagger$ a(b) stand for a$\times10^{\textrm{b}}$.}
   \label{initial_abundances}
   \begin{center}
   \begin{tabular}{lrr}
   \hline
   \hline
   Element			&	$n_i/n_\textrm{H}^\dagger$ 	&  References  \\
   \hline
   H$_2$ 			&	0.5		\\
   He				&	0.09  			&  1  \\	
   N				&	6.2(-5)			&  2  \\	
   O				&	2.4(-4)			&  3  \\	      
   C$^+$			&	1.7(-4)			&  2  \\	   
   S$^+$			&	8.0(-8)			&  4  \\	      
   Si$^+$			&	8.0(-9)			&  4  \\	
   Fe$^+$			&	3.0(-9)			&  4  \\
   Na$^+$			&	2.0(-9)			&  4  \\
   Mg$^+$			&	7.0(-9)			&  4  \\
   P$^+$			&	2.0(-10)  			&  4  \\
   Cl$^+$			&	1.0(-9) 			&  4  \\
   F				&	6.68(-9)			&  5  \\
   \hline
   \end{tabular}
   \end{center}
   \medskip
   (1) see discussion in \citet{Wakelam08}, (2) \citet{Jenkins09}, (3) see discussion in \citet{Hincelin11}, (4) low metal abundance from \citet{Graedel82}, (5) Depleted value from \citet{Neufeld05}.
\end{table}

\subsection{General effect of the 3-phase model on the ice composition}
\label{sec:general_effect}
In this section we discuss the impact of the introduced layered structure of the grain on the major ice compounds. We ran two types of models in which the mantle is considered as (1) chemically inert (labeled as model $a$), and (2) fully chemically active (i.e. including photodissociation, swapping and reactions ; labeled as model $b$). To study the effect of the reaction-diffusion competition, we ran these two models with and without this process. For comparison, we compare our results to the 2-phase model.

\begin{figure*}
\includegraphics[width=11.5cm,trim = 4cm 2.5cm 2cm 5cm, clip,angle=270]{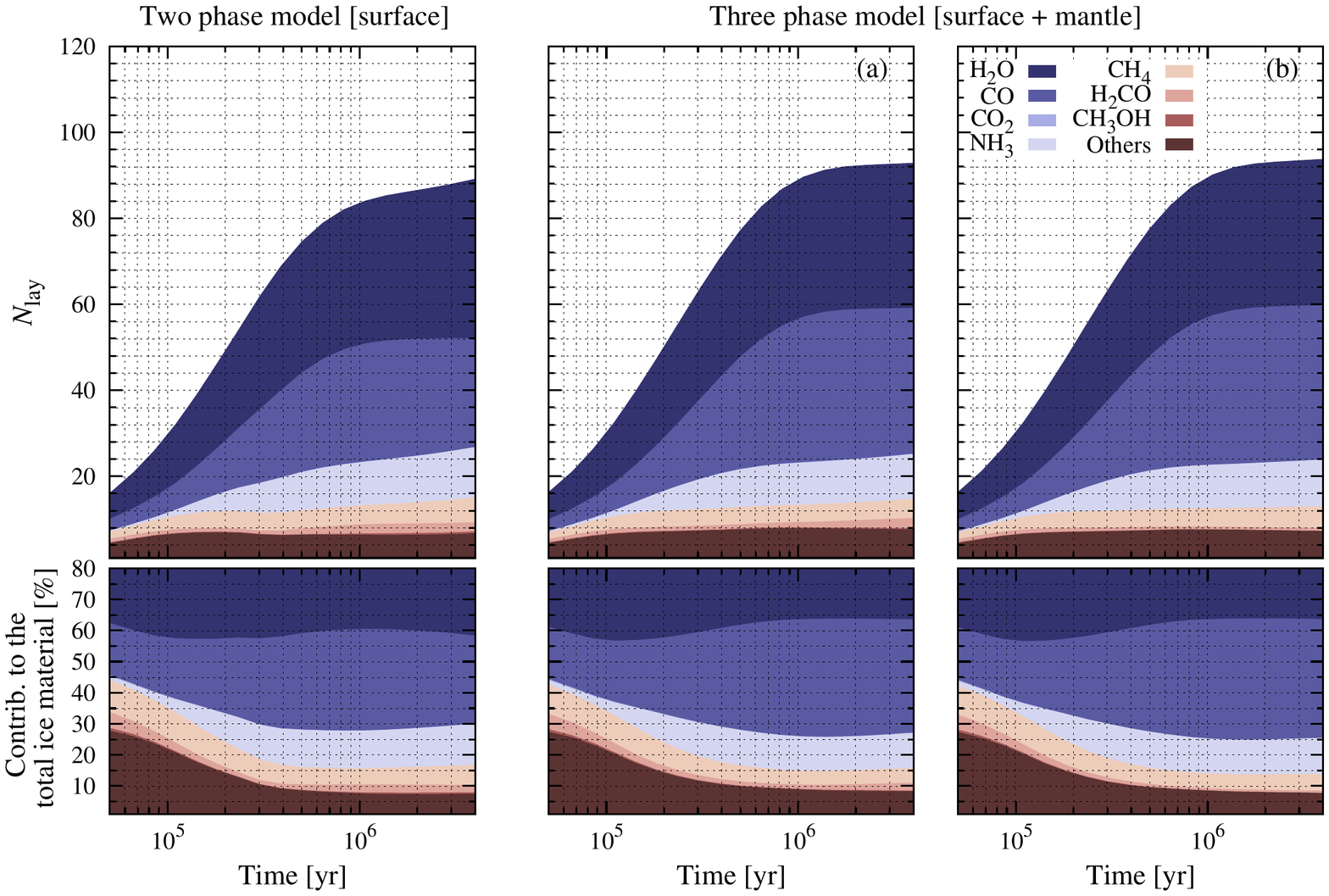}
\caption{Computed ice thickness and contribution to the total ice material for major ice compounds, when the reaction-diffusion competition is not activated, as a function of time. Left: Results obtained with the 2-phase model. Middle: Results obtained with the 3-phase model where the mantle is considered as inert (model $a$). Right: Full surface and mantle chemistry (model $b$). Note that CO$_2$, H$_2$CO and CH$_3$OH have very low abundances in these three models and may not be clearly visible on this plot.}
\label{Fig:ice_noreacdiff}
\end{figure*}

\begin{figure*}
\includegraphics[width=11.5cm,trim = 4cm 2.5cm 2cm 5cm, clip,angle=270]{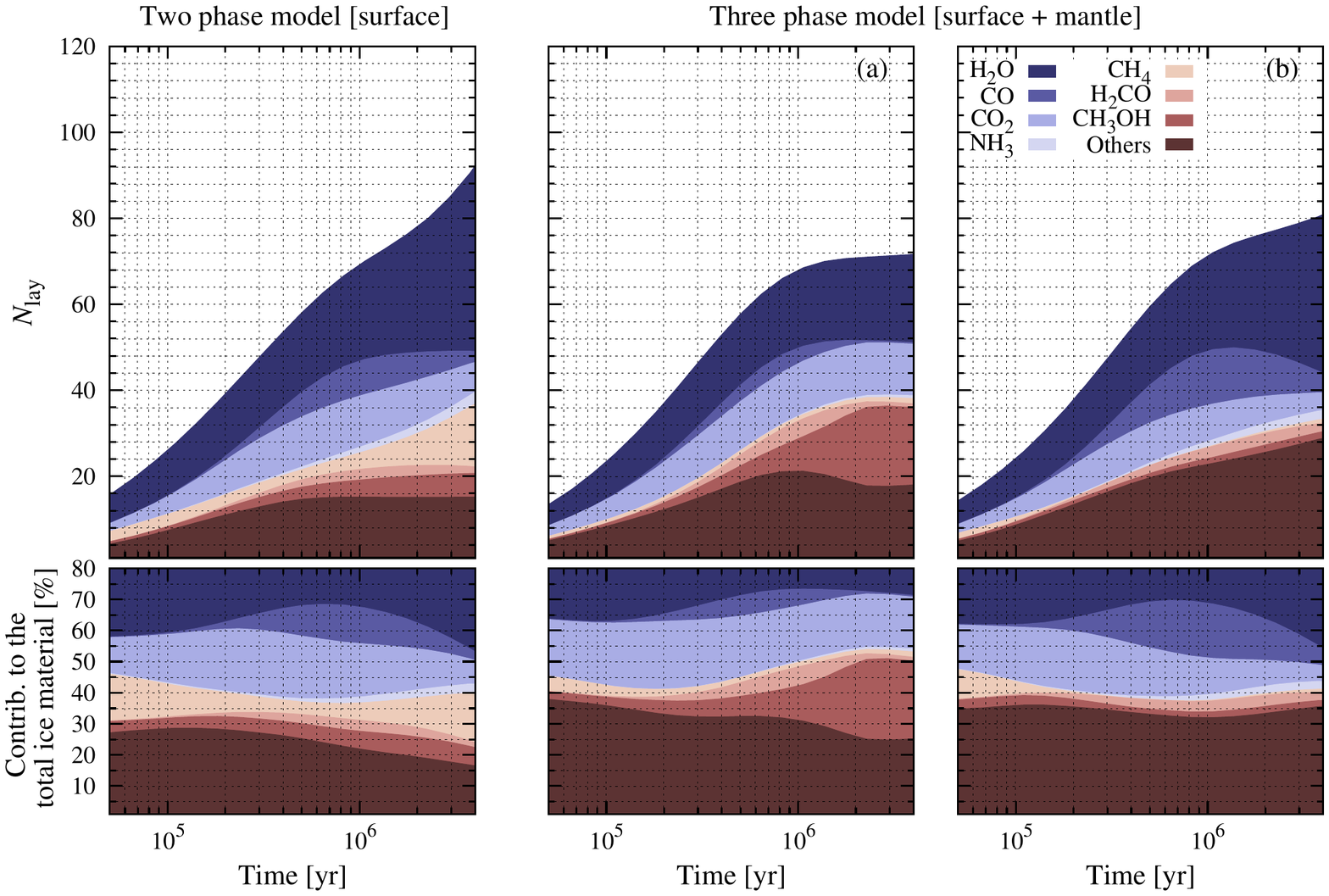}
\caption{Same as Fig. \ref{Fig:ice_noreacdiff} but with the incorporation of the reaction-diffusion competition. Note that NH$_3$, CH$_4$ and H$_2$CO have very low abundances in these three models and may not be clearly visible on this plot.}
\label{Fig:ice_reacdiff}
\end{figure*}

Figures \ref{Fig:ice_noreacdiff} and \ref{Fig:ice_reacdiff} show the model predictions for the major ice compounds between 0.5 and 4~Myr, with and without the reaction-diffusion competition respectively. In both cases, the left panel shows the results for 2-phase model, the middle one, the results for the 3-phase model where the mantle is considered as inert (model $a$) and the right panel the results when the full mantle chemistry is accounted (model $b$). In all the models, we show the observed main ice components: H$_2$O, CO, CO$_2$, NH$_3$, CH$_4$, H$_2$CO and CH$_3$OH for which the relative amounts of each molecule vary.

For all cases, a thickness of $\gtrsim80$ monolayers of ice is reached after 1~Myr. The computed ice is very water rich in all the simulations and H$_2$O represents at least $30\%$ of the total material of the ice (see the bottom panel of Fig. \ref{Fig:ice_noreacdiff} and Fig. \ref{Fig:ice_reacdiff}). In absence of reaction-diffusion competition, H$_2$O is mainly formed on the surface by the reaction H+OH$\rightarrow$H$_2$O. When this competition is taken into account, the main reaction path leading to H$_2$O ice is H$_2$+OH$\rightarrow$H$_2$O+H \citep[$E_\text{A}=2100$K;][]{Baulch84} on the surface rather than reaction with atomic H. This difference can be explained by the fact that when considering the reaction-diffusion competition, the rates of reaction with activation barriers are enhanced by several orders of magnitude (this point will be discussed in Section \ref{sec:gen_effects}). The same formation path was found by \citet{Cuppen07} using the CTRW Monte Carlo code technique and \citet{Garrod11} that also accounts for this competition.

If the reaction-diffusion is not taken into account, CO can be very abundant as well, and can account, as water ice, for almost $\sim30\%$ of the total ice material. When the reaction-diffusion competition is activated, the CO abundance is highly decreased due to the enhanced O+HCO reaction to form CO$_2$. In opposition to \citet{Garrod11}, the reaction with barrier OH+CO$\rightarrow$CO$_2$ + H \citep[$E_\text{A}=150$K;][]{Fulle96} is found to be inefficient compared to this reaction path and represents, in the best case, only few percents of the total flux leading to CO$_2$ formation. This may arise from the fact that in \citet{Garrod11} the barrier thickness $a$ used to calculate the square-well tunneling rate of the reaction H+CO$\rightarrow$HCO is set to $2\AA$ whereas we use a typical value of $1\AA$ for all the reactions that have a barrier as commonly assumed in most gas-grain models \citep{Tielens82,Hasegawa92}.

In both 2-phase models (Figs. \ref{Fig:ice_noreacdiff} and \ref{Fig:ice_reacdiff}), NH$_3$ and CH$_4$ are efficiently produced by the successive hydrogenation of nitrogen and carbon atoms on grains and can represent as much as $\sim 10\%$ of the total ice material. The fact that these two molecules  are produced with a slightly smaller abundance in the 3-phase models is mainly due to the fact that we use a low diffusion-to-binding ratio of 0.4 for the surface as compared to 0.5 for the 2-phase model. This implies that atoms (other than H) or light molecules move more rapidly on the surface by thermal hopping, making reactions involving such species more competitive in front of hydrogenation reactions.

In all models, there is a non negligible contribution coming from other molecules stated as "Others" on Figs. \ref{Fig:ice_noreacdiff} and \ref{Fig:ice_reacdiff}. The fraction of the ices represented by these species can be as high as 30\% at 2 Myr in models with the reaction-diffusion competition. In the model where the full chemistry is activated for the surface and the mantle (model $b$), this contribution comes principally from N-bearing species such as HCN, N$_2$, CH$_3$NH$_2$ and HC$_3$N, but also species such as CH$_3$OCH$_3$, HCOOH, HCO, CH$_3$O, CH$_3$CH$_2$OH, C$_2$H$_6$, CH$_2$OH, CH$_3$CCH and C$_2$H$_5$. The major ice compounds (i.e. species having an abundance larger than 0.2\% with respect to solid water) for this model are summarized in Table \ref{Tab:ice_mod}, for $t=2$Myr. From this Table, it can be seen that HCN, N$_2$, CH$_3$OCH$_3$ and HCOOH account for $\lesssim 5$ monolayers (denoted as ML hereafter) each of the total ice thickness, CH$_3$NH$_2$ and CH$_3$O for $\sim1$ML. Species like CH$_3$CH$_2$OH, C$_2$H$_6$, CH$_2$OH, CH$_3$CCH, C$_2$H$_5$ are also present and account for $0.6-0.7$ML each, i.e. abundances $\sim 2.0-2.5\%$ with respect to solid water. Such a high abundance of complex organic molecules (e.g. CH$_3$OCH$_3$) may not be realistic since no abstraction reactions are considered for the formation of such molecules. The fact that S-bearing species are absent from this list is due to the low sulfur initial abundance used, which comes from the low metal abundance of \citet{Graedel82}. As emphasized by \citet{Theule13}, a large part of these species have not been observed in interstellar ices but are suspected to be present in ices of molecular clouds.
Such high abundance of radicals, such as HCO and CH$_3$O, has already been reported by \citet{Taquet12} and \citet{Chang14}. Unlike \citet{Taquet12}, \citet{Chang14} considers an active mantle chemistry, i.e including mantle photodissociation, diffusion and reaction but using a unified microscopic-macroscopic Monte Carlo code. For these two species, the results obtained with our 3-phase model are found to be very similar to those presented in \citet{Chang14}. Their model also predicts a very high abundance of OH radical, which is not the case in our simulations. This difference can be due to the fact that they do not allow any H$_2$ to remain on the grains because of the rapid H$_2$ accretion and desorption which considerably slows down the Monte Carlo simulations. Consequently they have no reactions involving H$_2$ at the surface and in the mantle of the grains. In our case such high abundance of OH may not be observed due to the reaction H$_2$+OH$\rightarrow$H$_2$O+H \citep[$E_\text{A}=2100$K;][]{Baulch84}, which is found to be, as already mentioned, the major reaction leading to H$_2$O. Since in 2-phase models, the surface chemistry is treated assuming that the ice surrounding the interstellar grain core can be represented as an isotropic lattice with periodic potential (i.e. the surface and the mantle have same properties), this chemical diversity is not as important, because photodissociation is counterbalanced by hydrogenation reactions. In the case of 3-phase modelling, this "equilibrium" is broken by the difficulty for species to move in the ice. 

\begin{table}
   \caption{Calculated ice abundances at $t=2$Myr$^\dagger$ using the developed 3-phase model. In this model, the mantle is considered as active (model $b$) and the diffusion-to-binding ratio was set to $E^s_\text{diff}/E_\text{des}=0.4$.}
   \label{Tab:ice_mod}
   \begin{center}
   \begin{tabular}{lrrr}
   \hline
   \hline
   Molecule	&	$n(x)/n_{\text{H}_2\text{O}}$& $n(x)/n_\text{H}$	& 	$N_\text{ML}$\\
   			&	[$\%$]				&	[$\times 10^{-6}$]	&		\\
   \hline
   \multicolumn{4}{c}{Surface + Mantle}\\
   \hline
   H$_2$O$^a$ 			&	100.0&    97.5	&	28.8	\\
   \hline
   \multicolumn{4}{c}{Surface}\\
   \hline
   H$_2$O$^a$			&	2.7	&	2.6	&	0.8	\\
   CO	$^a$				&	0.8	&	0.8	&	0.3	\\
   CO$_2$$^a$			&	0.5	&	0.5	&	0.2	\\
   HCN$^c$			&	0.5	&	0.5	&	0.1	\\
   N$_2$$^b$			&	0.4	&	0.4	&	0.1	\\
   HCOOH$^b$ 			&	0.3	&	0.3	&	0.08	\\
   HCO				&	0.2	&	0.2	&	0.06	\\
   NH$_3$$^a$			&	0.2	&	0.2	&	0.04	\\
   CH$_3$OH$^a$		&	0.2	&	0.2	&	0.04	\\
   \hline
   \multicolumn{4}{c}{Mantle}\\
   \hline
   H$_2$O	$^a$			&	97.2	&      94.8	&	28.0	\\
   CO$^a$				&	33.0	&	32.2	&	9.5	\\
   CO$_2$$^a$			&	22.5	&	20.0	&	5.9	\\
   HCN$^c$			&	19.0	&	18.6	&	5.5	\\
   N$_2$$^b$			&	14.6	&	14.2	&	4.2	\\
   CH$_3$OCH$_3$ 		&	13.0&	12.6	&	3.7	\\
   HCOOH$^b$			&	11.0	&	10.5	&	3.1	\\
   H$_2$CO$^b$		&	9.0	&	8.9	&	2.6	\\
   HCO				&	7.5	&	7.3	&	2.1	\\
   NH$_3$$^a$			&	5.9	&	5.7	&	1.7	\\
   CH$_3$OH$^a$		&	5.7	&	5.6	&	1.7	\\
   CH$_3$NH$_2$$^c$	&	5.1	&	5.0	&	1.5	\\
   CH$_3$O			&	4.1	&	4.0	&	1.2	\\
   CH$_3$CH$_2$OH$^b$&	2.6	&	2.5	&	0.7	\\
   C$_2$H$_6$$^b$		&	2.5	&	2.4	&	0.7	\\
   CH$_2$OH			&	2.4	&	2.3	&	0.7	\\
   CH$_3$CCH			&	2.1	&	2.0	&	0.6	\\
   CH$_4$$^a$			&	1.7	&	1.6	&	0.5	\\
   C$_2$H$_5$			&	1.4	&	1.3	&	0.4	\\
   CH$_3$CHO$^b$		&	0.6	&	0.6	&	0.2	\\
   CH$_3$OCH$_2$		&	0.6	&	0.6	&	0.2	\\
   NO$^c$				&	0.5	&	0.5	&	0.1	\\	
   HC$_3$N			&	0.4	&	0.4	&	0.1	\\
   HNO				&	0.2	&	0.2	&	0.07	\\
   C$_2$H$_4$$^c$		&	0.2	&	0.2	&	0.05	\\   
   \hline
   \end{tabular}
   \end{center}
   \medskip
   The top panel shows the total solid water abundance and its contribution to the total ice material in term of number of monolayers, i.e. denoted as $N_\text{ML}$. The second panel is related to the major surface compounds and the third panel is related to the mantle species.\\
   $^\dagger$ Best time defined to fit observations in MYSOs, LYSOs and toward background stars. See Section \ref{sec:ice_comp}.\\
   $^a$Molecules securely identified \citep{Boogert15}.\\
   $^b$Likely and possibly identified species as well as upper limit detections \citep{Boogert15}.\\
   $^c$Species that are suspected to be present \citep{Theule13}.
\end{table}

The fact to include or not the full chemistry in the mantle strongly affects the results. In model $b$ with the reaction-diffusion competition, the reactions on the surface layers lead mostly to large molecules such as CH$_3$OH and CH$_3$NH$_2$ because of the faster diffusion of heavy species. These complex species are then incorporated in the mantle (when there is accretion of other gas-phase species) where they are dissociated by cosmic-ray induced photons. The radicals resulting from the dissociations (e.g. NH$_2$, CH$_3$) are then mostly hydrogenated into more simple molecules such as NH$_3$ and CH$_4$. In the 3-phase model $a$ with the reaction-diffusion competition and an inert mantle (middle panel of Fig.~\ref{Fig:ice_reacdiff}), CO, NH$_3$ and CH$_4$ abundances are very small compared to observations in interstellar ices because CO has been transformed into CO$_2$ while nitrogen and carbon are locked in more complex species in the chemically inactive mantle (e.g. CH$_3$OH).
 In Model $b$ with the reaction-diffusion competition and an active mantle, NH$_3$ is formed in the grain mantle by the reaction H$_2$+NH$_2$$\rightarrow$NH$_3$+H \citep[$E_\text{A}=6300$K;][]{Mitchell84} and by hydrogenation of NH$_2$, resulting from the dissociation of CH$_3$NH$_2$. 
Similarly, CH$_4$ is formed in the grain mantle by photodissociation of CH$_3$OCH$_3$ and by hydrogenation of CH$_3$ which is produced in the mantle by photodissociation of CH$_3$OH and CH$_3$NH$_2$.

The chemistry we presented depends on the efficiency of the accretion process compared to the diffusion time on the surface. If the diffusion time of species on the surface is  larger than the accretion time scale (and so the transfer of species from the surface to the mantle where they can diffuse less), then the formation of complex species would be much less efficient. In fact, at the densities of cold cores the continuous renewal of the surfaces by fresh material accreting from the gas-phase does not impact the surface reactivity by transfer of species from the surface to the mantle. The typical time for a species to arrive on a particular grain of radius $r_\text{dust}$ is given by:

\begin{equation*}
\begin{split}
\tau_\text{acc}^{-1}(i) &= n(i)v(i)\pi r_\text{dust}^2 \\
&\simeq 8\times10^2 \bigg(\frac{r_\text{dust}}{0.1\mu\text{m}}\bigg) \bigg(\frac{\text{T}_\text{gaz}}{10\text{K}}\bigg)^{1/2}\bigg(\frac{\text{m}_i}{1.0~\text{amu}}\bigg)^{-1/2}\text{n}(i)~[\text{yr}^{-1}]
\end{split}
\end{equation*}
If we consider that the initial concentration of the most abundant species (except H and H$_2$) under dark cloud conditions (i.e. at n$_\text{H}=2\times10^4$\cmt) is $\simeq 1$ molecule.\cmt, it means that a grain accretes $\simeq10^2$ molecules per year. It therefore requires $\simeq 10^4$ yr to cover $10^6$ and so on. The density required to see an effect of the renewal of the surface by successive accretion can be obtained by comparing this time to the typical diffusion time of a species on the surface $t^s_\text{diff}$. At n$_\text{H}=2\times10^4$\cmt~and using a diffusion-to-binding ratio of 0.4, the diffusion time of light atoms, e.g. H ($E^s_\text{diff}=230$K), O ($E^s_\text{diff}=320$K), N ($E^s_\text{diff}=320$K), etc, or molecules such as H$_2$ ($E^s_\text{diff}=180$K), O$_2$ ($E^s_\text{diff}=400$K) or CO ($E^s_\text{diff}=460$K)..., is shorter than the typical time to cover one layer by successive accretion where  $t^s_\text{diff}(\text{H}_2)\simeq10^{-6}$ yr, $t^s_\text{diff}(\text{H})\simeq10^{-4}$ yr, $t^s_\text{diff}(\text{O})\simeq$ $t^s_\text{diff}(\text{N})\simeq 1$ yr, $t^s_\text{diff}(\text{O}_2)\simeq 10^4$ yr and $t^s_\text{diff}(\text{CO})\simeq 10^6$ yr. It results that for O and N, for example, this effect can be important for $n_\text{H} \gtrsim 10^{8}$\cmt.
In the case of the formation of CO$_2$, since $E_\text{diff}^s(\text{HCO}) > E_\text{diff}^s(\text{O})$, for densities $n_\text{H} \gtrsim 10^{8}$\cmt, the surface reaction O+HCO will start to become inefficient due to the rapid transfer of both reactants into the icy mantle.

\subsection{Comparison with interstellar ice}
\label{sec:ice_comp}
 
We have compared the results of our 2-phase model and 3-phase models $b$ (both with the reaction-diffusion competition process and an active mantle) with the abundances of ices observed in the envelopes around young stellar objects, i.e. low-mass and massive young stellar objects (LYSOs and MYSOs respectively), and toward background stars (BG stars) reported by \citet{Boogert15}. For these models, we have determined the time of best fit \citep[following the method described in][]{Garrod07}. This time together with the modeled and observed abundances are listed in Table~\ref{Tab:ice_obs}. Two additional models $b$ were tested in which $E^s_\text{diff}/E_\text{des}=0.3$ and 0.5 and will be discussed in the following and in Section~\ref{sec:gas_modelisation}. 

While discussing the comparison, it is important to keep in mind that the observed ices toward such sources are known to be slightly to strongly affected by thermal processing \citep{Boogert15} and that such comparison may not be relevant for simulations run under dark cloud conditions, i.e. where the temperature is as low as 10K and the density is kept fixed. For this reason, our comparison is made over the entire range of observational values rather than averaged values and aims to estimate how our modelling results can qualitatively reproduce the observed ice composition in these sources. We do not show the results for models where the reaction-diffusion competition is not taken into account since we are not able to reproduce the observed CO$_2$ ice abundance and we strongly overestimate the CO abundance.

The results obtained with the 2-phase model show a good agreement with the observed abundances in ices except for CH$_3$OH and CH$_4$ which are  overproduced. In comparison, the 3-phase model calculation where the full mantle chemistry is considered and where $E^s_\text{diff}/E_\text{des}=0.4$ is assumed, shows a very good agreement with observations for all the observed species. As for the 2-phase model, the results obtained with the 3-phase model where we assumed $E^s_\text{diff}/E_\text{des}=0.5$ slightly overproduce CH$_3$OH and CH$_4$. The model with  $E^s_\text{diff}/E_\text{des}=0.3$ show a good agreement for these two molecules but underestimate the CO abundance and strongly overestimate the CO$_2$ abundance.

\begin{table*}
   \caption{Best fit ice composition against observation in MYSOs, LYSOs and toward BG Stars. All the models where run with the reaction-diffusion competition accounted.}
   \label{Tab:ice_obs}
   \centering
   \begin{tabular}{lcrrrrrrrrr}
   \hline
   \hline
   Model									& 	Best fit time [yr] 	&H$_2$O	&CO			&CO$_2$		&CH$_3$OH	&NH$_3$	&CH$_4$	& H$_2$CO\\
   \hline
   2-phase									& $2.0\times10^6$		&100		&25			&39			&17			&7		&24		&8	\\
   3-phase ($b$)~($E^s_\text{diff}/E_\text{des}=0.3$)	& $1.0\times10^6$		&100		&8			&90			&4			&5		&1		&4	\\
   3-phase ($b$)~($E^s_\text{diff}/E_\text{des}=0.4$) & $2.0\times10^6$		&100		&34			&23			&6			&6		&2		&9	\\
   3-phase ($b$)~($E^s_\text{diff}/E_\text{des}=0.5$) & $2.0\times10^6$		&100		&43			&35			&7			&9		&18		&11	\\
   \hline
   Observations$^a$\\
   \hline
   BG Stars	&	&100		&20-43		&18-39		&6-10		&<7	&<3			&			\\
   MYSOs		&	&100		&4-14		&12-25		&5-23		&$\sim$7	&1-3		&$\sim$2-7	\\
   LYSOs		&	&100		&12-35		&23-37		&5-12		&4-8	&3-6			&$\sim$6		\\
   \hline
   \multicolumn{8}{l}{$^a$Observed ice composition taken from the review of \citet{Boogert15}.}
   \end{tabular}
\end{table*}

\begin{figure*}
\includegraphics[width=7.5cm,trim = 1cm 2cm 9.5cm 1cm, clip,angle=270]{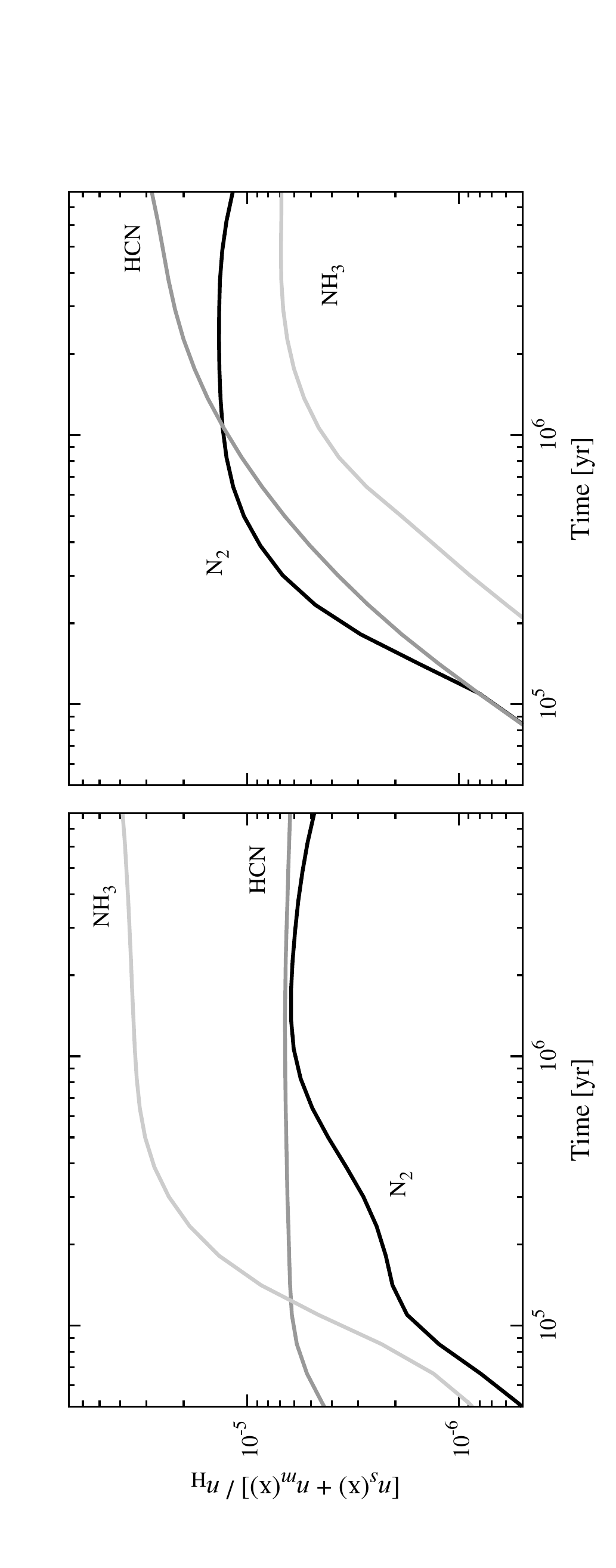}
\caption{HCN, N$_2$ and NH$_3$ abundances on grains with respect to $n_\text{H}$ as a function of time obtained with the 3-phase model $b$. The left panel corresponds to the results obtained without taking into account the reaction-diffusion competition, while the right panel corresponds to the results when this competition was accounted.}
\label{Fig:NbearingSpecies}
\end{figure*}

As seen in Table \ref{Tab:ice_mod}, the reservoir of O-bearing species is water while CO, CO$_2$ contain most of the carbon. Surprisingly, HCN and N$_2$ are found to be the reservoirs of nitrogen in the icy mantle of the grain, i.e. more abundant than NH$_3$. Fig. \ref{Fig:NbearingSpecies} presents the calculated abundances of N$_2$, HCN and NH$_3$ on the grain with the 3-phase model ($b$) where $E^s_\text{diff}/E_\text{des}=0.4$. In this figure, the left panel corresponds to the results obtained without the reaction-diffusion competition activated and the right panel when this competition was accounted. When the reaction-diffusion competition is not taken into account, at relevant ages (i.e. $t \gtrsim 10^5$ yrs), NH$_3$ on the grains is always the major nitrogen bearing species. When the reaction-competition is taken into account, the abundance of NH$_3$ is highly reduced due to a lower abundance of H. This lower abundance of H is caused by the enhanced reaction rates that have a barrier and that consume the atomic hydrogen. On the other hand, the production of HCN on the grain is enhanced, compared to the case where the reaction-diffusion competition is not taken into account, by the reaction H$_2$+CN$\rightarrow$HCN+H \citep[$E_\text{A}=2070$K;][]{Mitchell84}. 

Since comets could reflect (at least partially) the primordial material that formed them \citep{Oberg11,Boogert15}, it is interesting to compare qualitatively our results with observations in these objects. We focus on the comet 67P/Churyumov-Gerasimenko for which in situ measurements were acquired, even though there seems to be a large diversity on the composition of comets. The ROSINA experiment on board ROSETTA spacecraft, has found that in the coma of the comet 67P/Churyumov-Gerasimenko, HCN would be more abundant than NH$_3$ \citep{LeRoy15} and would in fact be more abundant than N$_2$ \citep{Rubin15}. HC$_3$N was also detected in the coma of the comet and is predicted by our model. CH$_3$NH$_2$ however is the fourth most abundant N-bearing species in our model and was apparently not detected in the coma of the comet 67P/Churyumov-Gerasimenko but seems to be present on the comet as seen by the COSAC experiment onboard of the Rosetta's lander Philae  with an abundance of 0.6 relative to water \citep{Goesmann15}. If we make the same comparison with other volatiles identified in this comet by \citet{LeRoy15}, we find the same species in our most abundant ones, except for C$_2$H$_2$.

\section{Effect on the gas-phase abundances}
\label{sec:gas_modelisation}

In this section we focus on the gas-phase abundances only and study the impact of the 3-phase model on the predicted gas-phase abundances. We first discuss the differences in the gas-phase predictions of the 3-phase and 2-phase models and then look at the impact on the comparison with observed abundances in the well known TMC-1 (CP) and L134N dark clouds. For this section, we can consider either the model $a$ or $b$ since the chemistry or its absence in the mantle does not influence the gas-phase composition; i.e. because net accretion from the gas-phase dominates.
In order to see the impact of varying the diffusion-to-binding energy ratio on the modeled gas-phase abundances, we ran our 3-phase model using $E^s_\text{diff}/E_\text{des}=0.3$, 0.4 and 0.5.

\subsection{General effects}
\label{sec:gen_effects}

\begin{figure*}
\includegraphics[width=8.3cm,trim = 2cm 2cm 7cm 2cm, clip,angle=270]{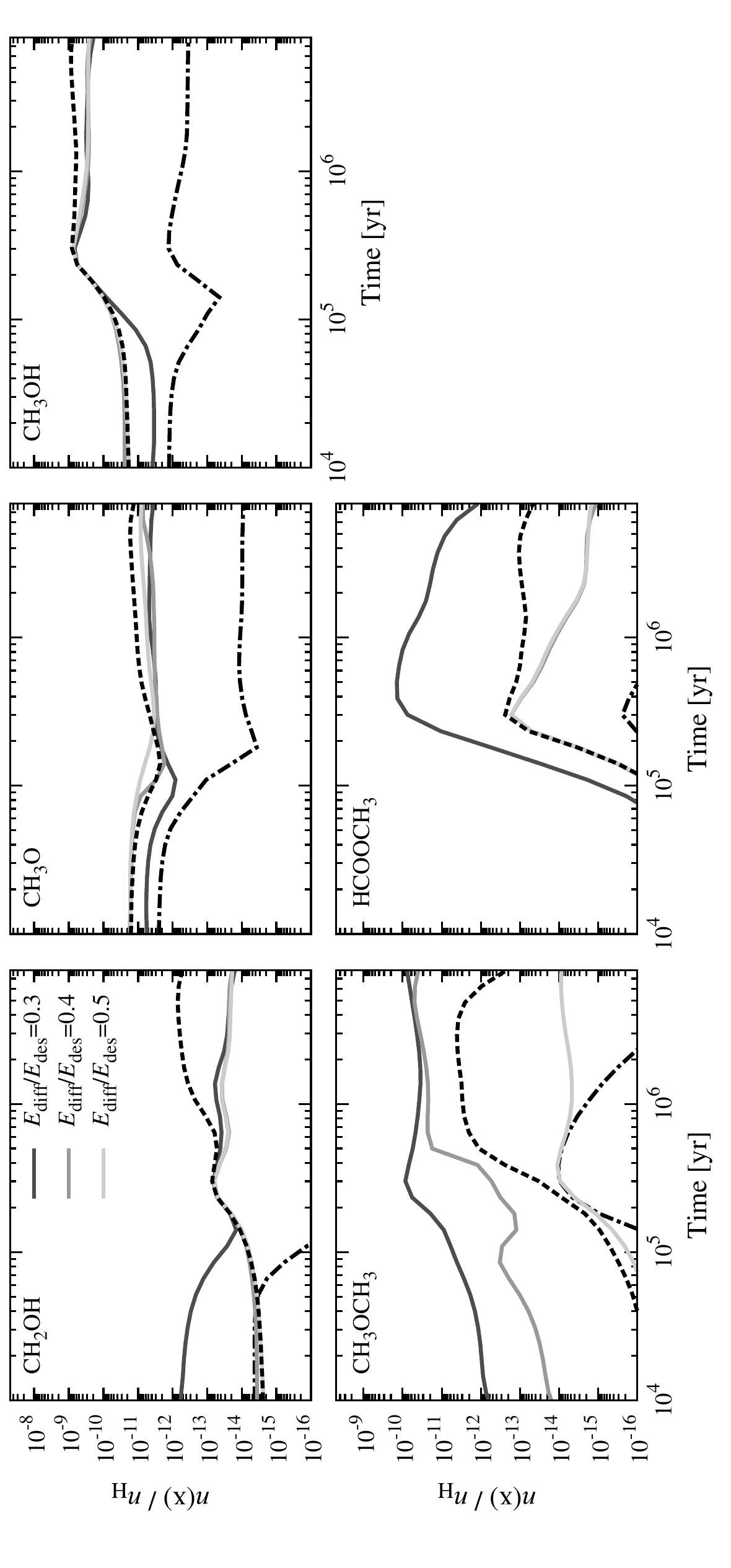}
\caption{Abundance with respect to $n_\text{H}$ of selected gas-phase species as a function of time. The black dashed lines corresponds to the results obtained with the 2-phase model where $E_\text{diff}/E_\text{des}=0.5$. The gray solid lines correspond to the results obtained with the 3-phase model with $E^s_\text{diff}/E_\text{des}=0.3$, 0.4 and 0.5. All these results, except for the black dash-dotted lines, were obtained by accounting for the reaction-diffusion competition. The black dash-dotted lines corresponds to the result obtained with the 3-phase model with $E_\text{diff}/E_\text{des}=0.4$ but without the reaction-diffusion competition.}
\label{Fig:reac-diff_comp}
\end{figure*}

As seen in Section \ref{sec:ice_comp}, 3-phase modelling, where the reaction-diffusion competition was not activated, produces worse results than when accounted, especially for CO$_2$ on grains that is highly underestimated. Nevertheless, it is interesting to see its impact on the computed gas-phase abundances. Figure \ref{Fig:reac-diff_comp} shows the calculated abundance of five selected gas-phase species as a function of time. The black dash-dotted lines corresponds to the results where the reaction-diffusion competition was not taken into account and $E^s_\text{diff}/E_\text{des}=0.4$ was assumed. The gray solid lines show the results when this competition was accounted and where $E^s_\text{diff}/E_\text{des}=0.3$, 0.4 and 0.5 respectively. The black dashed lines show the results obtained with the 2-phase model where we used $E_\text{diff}/E_\text{des}=0.5$. As presented here the reaction-diffusion competition on the grain surface can have a strong impact on the modeled gas-phase abundances.  For models where $E^s_\text{diff}/E_\text{des}=0.4$, all those species are formed at the surface of the grains and released in the gas-phase at their formation by chemical desorption. CH$_3$O, CH$_2$OH and CH$_3$OH are formed via the successive hydrogenations of CO, which present two activation barriers for reactions H+CO and H+H$_2$CO on the order of $ E_\text{A}\simeq 10^3$ K. CH$_3$OCH3 and HCOOCH$_3$ are formed by the successive hydrogenation of CH$_3$CHO (which is formed by CH$_3$+HCO) and by HCO+CH$_3$O respectively.

For all these species, the predicted gas-phase abundances are larger when the reaction-diffusion competition is taken into account (dark gray solid lines versus black dash-dotted lines on Fig. \ref{Fig:reac-diff_comp}). This is due to a strong enhancement of activation-energy-barrier-mediated reactions when this competition is considered. Since this competition is introduced to take into account the fact that two molecules mobile on the surface may meet in the same binding sites and possibly react before one of the reactant hops into an adjacent binding site, this result is not straightforward considering the high mobility of hydrogen atoms. This comes from the manner that we treat (or not) the reaction-diffusion competition. In our case, we always assume that hydrogenation reactions that have a barrier proceed through tunneling of H atoms. Considering a barrier thickness of 1$\AA$, the probability of reaction without taking into account the reaction-diffusion competition is $\kappa_{ij}^*=\exp[-2(a/ \hbar)(2\mu E_{\text{A},ij})^{1/2}] \sim 10^{-9}$ where $i$ stands for H and $j$ for CO or H$_2$CO for example. When the reaction competition is activated, $\kappa_{ij} \simeq \nu_{ij} \kappa_{ij}^*/ [\nu_{ij} \kappa_{ij}^* + k_\text{hop}^s(\text{H})]$, considering that only H is moving and that $k_\text{hop}^s(\text{H}) \gg k_\text{des}(\text{H})$. At T$_\text{dust}=10$K, with $E_\text{diff}^s(\text{H})=230$K and considering $\nu_{ij} \simeq \nu_\text{H}$ of the order of $10^{12}$ s$^{-1}$, $k_\text{hop}^s(\text{H})=\nu_{\text{H}} \exp(-E_\text{diff}^s(\text{H})/\text{T}_\text{dust})\sim 10^2$ s$^{-1}$. This leads to $\kappa_{ij} = 0.90$, a value $\sim 10^9$ times higher than $\kappa_{ij}^*$, so the reaction probability is greatly enhanced considering the reaction-diffusion competition although H is very mobile at the grain surface. Increasing the diffusion-to-binding energy ratio (except for H that we consider known from theory) does not modify the calculated abundances of CH$_2$OH, CH$_3$O and CH$_3$OH since $\kappa_{ij}$ depends mostly on $k_\text{hop}^s(\text{H})$. On the contrary, the calculated abundances of CH$_3$OCH$_3$ and HCOOCH$_3$ increase when $E^s_\text{diff}/E_\text{des}$ decreases because these species are formed from reactions between radicals rather than just hydrogenation. 

\subsection{Comparison with TMC-1 and L134N dark clouds}

\begin{figure*}
\includegraphics[width=7.5cm,trim = 1cm 2cm 9.5cm 1cm, clip,angle=270]{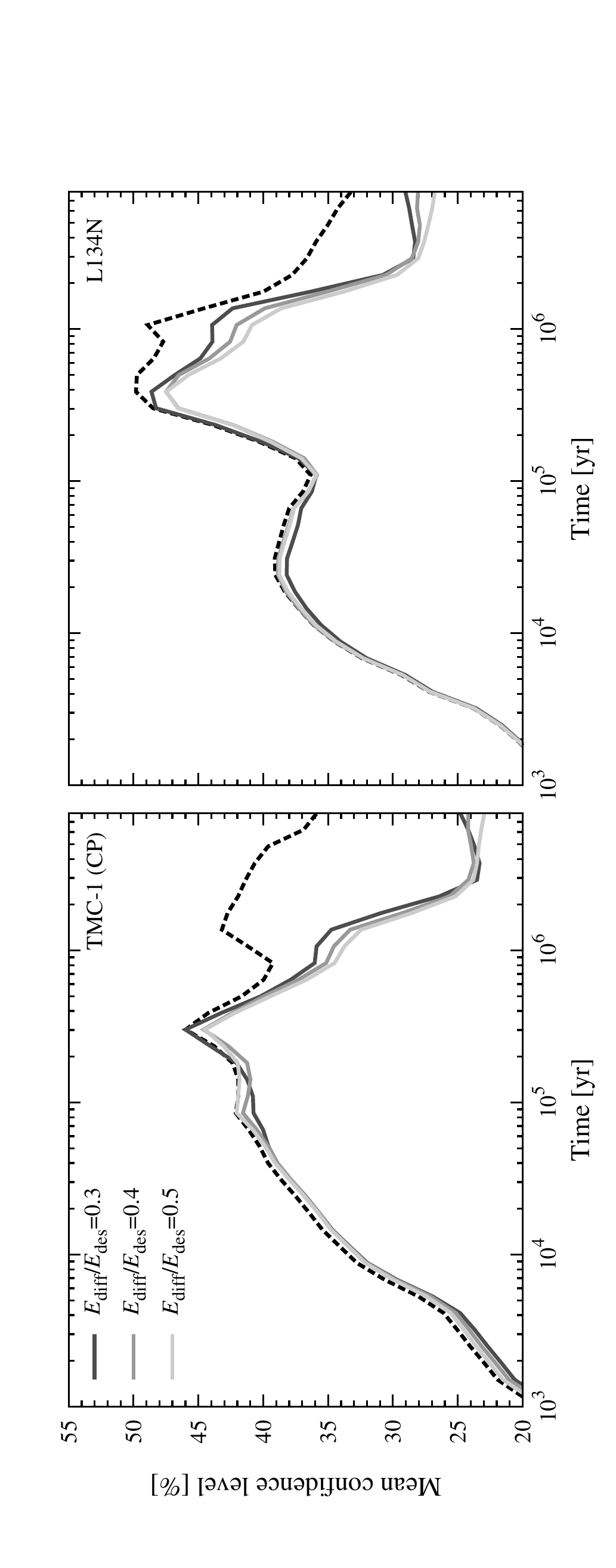}
\caption{Calculated mean confidence level as a function of time for fit with TMC-1 (CP) (left panel) and L134N (right panel) observations. The dashed lines corresponds to the results obtained with the 3-phase model $b$ where $E^s_\text{diff}/E_\text{des}=0.3$, 0.4 and 0.5. The solid lines corresponds to the results obtained using the 2-phase model with $E_\text{diff}/E_\text{des}=0.5$. In all cases, the reaction-diffusion competition was accounted.}
\label{Fig:distance}
\end{figure*}

We compare the gas-phase predicted abundances with our 3-phase model including the reaction-diffusion competition and the full chemistry in the mantle (with $E^s_\text{diff}/E_\text{des}=0.3$, 0.4 and 0.5) to the molecules observed in the two dark clouds TMC-1 (CP) and L134N listed in  \citet{Agundez13}. As seen in Section \ref{sec:ice_comp}, it is important to note that such comparison should be taken with care since these sources are known to have a complex structure \citep[see][for L134N for example]{Pagani07} and could have different physical history. Fig. \ref{Fig:distance} presents the mean confidence level calculated following \citet{Garrod07} for fit with TMC-1 (CP) and L134N observations. As it can be seen, the new model proposed here does not have a strong impact on the general confidence level which lies between $\sim 45 \%$ for TMC-1 (CP) and $\sim 50 \%$ for L134N at its maximum. On average, this corresponds to less than a factor of 10 deviation between modeled and observed abundances for the best matches. The fact that the mean confidence level does not vary significantly from 2-phase to 3-phase modelling is mainly due to the fact that most of the molecules observed in both, TMC-1 (CP) and L134N clouds, originates from gas phase chemistry. The 3-phase model where $E^s_\text{diff}/E_\text{des}=0.3$ was assumed gives a better mean confidence level compared to the results obtained with $E^s_\text{diff}/E_\text{des}=0.4$ and 0.5. However, as seen in Section \ref{sec:ice_comp}, the model where $E^s_\text{diff}/E_\text{des}=0.3$ is assumed gives a worse agreement for the comparison with ice observations in comparison to the results obtained with $E^s_\text{diff}/E_\text{des}=0.4$. The use of the 3-phase model suggests a chemical-age for both TM1-CP and L134N dark clouds of few $10^5$ yr whereas the 2-phase model proposes two ages with the the same confidence level of a few $10^5$ yrs and $\sim 10^6$ yr. This later age is strongly rejected by the use of the 3-phase model precisely due to large accretion on dust which is not counterbalanced by the desorption. It is however important to keep in mind that these chemical ages depends greatly on the model as well as on the comparison method used and can vary from one study to the other depending on the assumptions made. At the time of best agreement, for both 2-phase and 3-phase models, $\sim 60\%$ and $\sim 70 \%$ of the observed species in TMC-1 (CP) and L134N respectively are reproduced by the model within a factor of 10. For TMC-1 (CP), this represent 44 over 64 molecules observed and 36 over 50 for L134N.

\section{Conclusion}
\label{sec:conclusion}

In this paper we present a new 3-phase model for gas-grain simulations in which the mantle as well as the surface are considered as active. Mantle photodissociation, diffusion and reactions are taken into account.  Following \citet{Fayolle11}, to account for surface roughness, we consider that the two outermost monolayers are part of the surface. We also assume that the bulk diffusion is driven by the diffusion of water molecules in the ice as suggested by molecular dynamics simulations and experiments \citep{Ghesquiere15}. The diffusion-to-binding ratio of surface species is taken to be equal to 0.4, except for H for which we use a specific value coming from theoretical calculations \citep{AlHalabi07}. The proposed model successfully reproduces the TPD data on ice desorption and deals with zeroth- and first-order desorption kinetics.
The main conclusions of this study are the following:
\begin{enumerate}
\item Considering the reaction-diffusion competition or not has a strong impact on the modelling results. This competition seems indispensable to form CO$_2$ on grains as already found by \citet{Garrod11}. The reaction-diffusion competition not only affects the CO$_2$ abundance on grains but more generally all the species that are formed via reactions that have a barrier. In this context, H$_2$O ice is found to be formed by H$_2$+OH$\rightarrow$H$_2$O+H \citep[$E_\text{A}=2100$K;][]{Baulch84} on the surface rather than reaction with H \citep[in agreement with][]{Cuppen07}. NH$_3$ is found to be formed by H$_2$+NH$_2$$\rightarrow$NH$_3$+H \citep[$E_\text{A}=6300$K;][]{Mitchell84} and by successive hydrogenation of NH. HCN is predominantly formed by H$_2$+CN$\rightarrow$HCN+H \citep[$E_\text{A}=2070$K;][]{Mitchell84} in the grain mantle. By incorporating the reaction-diffusion competition, it becomes crucial to have a good approach to determine the abundance of H$_2$ in interstellar ices.
\item The low diffusion-to-binding energy ratio, used for surface, makes reactions involving light atoms or molecules other than H more competitive in front of hydrogenation reactions. This leads to a lower abundance of species formed by hydrogenation chains on the grains as well as in the gas phase.
\item The proposed 3-phase model qualitatively reproduces most of the observed ice species in envelopes of young stellar objects and toward background stars. A number of others species are found to be abundant on grains, with abundances of the order of few $\times 10^{-6}$ with respect to the total proton density. These species are essentially N-bearing species such as N$_2$, HCN and CH$_3$NH$_2$. Species like CH$_3$OCH$_3$, HCOOH, HCO, CH$_3$O, CH$_3$CH$_2$OH, C$_2$H$_6$, CH$_2$OH, CH$_3$CCH and C$_2$H$_5$ are also found to be abundant on the grain surface, i.e. with abundances $\gtrsim 10^{-6}$ with respect to the total proton density.
\item It is found that accounting for the reaction-diffusion competition modifies the major N-bearing species on the grain which go from NH$_3$ to N$_2$ and HCN. The abundance of NH$_3$ is found to be reduced due to a lower abundance of H caused by the reaction-diffusion competition. On the other hand, this competition is found to promote the formation of HCN by H$_2$+CN$\rightarrow$HCN+H \citep[$E_\text{A}=2070$K;][]{Mitchell84}.
\item The developed 3-phase model does not have a strong impact on the computed abundances of the observed species in TMC-1 (CP) and L134N dark clouds compared to the original 2-phase model before $10^6$ yrs. After $10^6$ yrs, we find that the agreement dramatically falls down compared to the agreement obtained with the 2-phase model. This  is the result of a dramatic fall down of the gas phase abundances due to the large accretion on dust which is, unlike the 2-phase model, not counterbalanced by the desorption. This strongly constrains the chemical-age of these clouds to be of the order of a few $10^5$ yrs and strongly rejects the largest ages of few $10^6$ yrs.
\end{enumerate}

\section*{Acknowledgements}
This work has been founded by the European Research Council (Starting Grant 3DICE, grant agreement 336474). The authors are also grateful to the CNRS program "Physique et Chimie du Milieu Interstellaire" (PCMI) for partial funding of their work. 

%
\bibliographystyle{mnras}
\bibliography{nautilus_3-phase} 

%

\appendix

\section{TPD experiments}
\label{sec:tpd_expe}
Desorption is a two step process in which first and zero order desorption kinetics are at work. The first order desorption is caused by the desorption of the volatile species from the surface while the zero order is caused by the replacement of desorbed surface species by mantle species. This last point is also known as the co-desorption process when the ice is composed of volatiles trapped in a more refractory ice that desorb with the refractory material.

One of the major problem of the 2-phase model is that the desorption does not take into account the zeroth-order kinetic desorption which leads to an unphysical treatment when net desorption occurs. In our model, this problem is solved by treating the surface and the mantle as separate phases and that the diffusion in the mantle is driven by the self diffusion of water in the ice, i.e. following the recent findings of \citet{Ghesquiere15}.

\begin{figure}
\includegraphics[width=7cm,trim = 2cm 3cm 2cm 6cm, clip,angle=270]{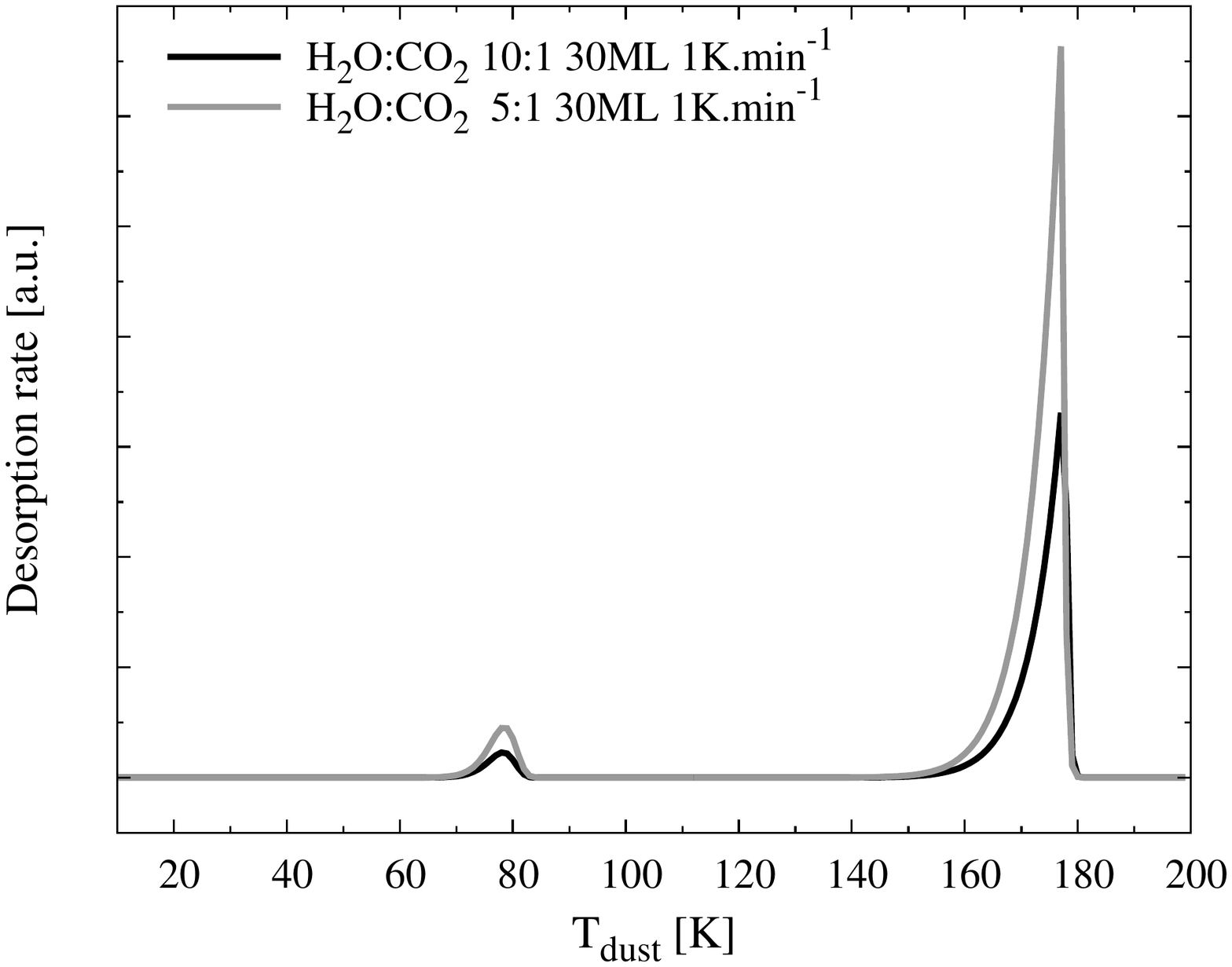}
\caption{Simulated CO$_2$ desorption rates as a function of T$_\text{dust}$ from two binary water dominated ice mixture. The heating rate used is 1K.min$^{-1}$. The black line corresponds to a 30ML ice composed of H$_2$O:CO$_2$ 10:1 and the gray line to a 30ML ice composed of H$_2$O:CO$_2$ 5:1.}
\label{Fig:TPD_h2o-co2_30ML}
\end{figure}

\begin{figure}
\includegraphics[width=7cm,trim = 2cm 3cm 2cm 6cm, clip,angle=270]{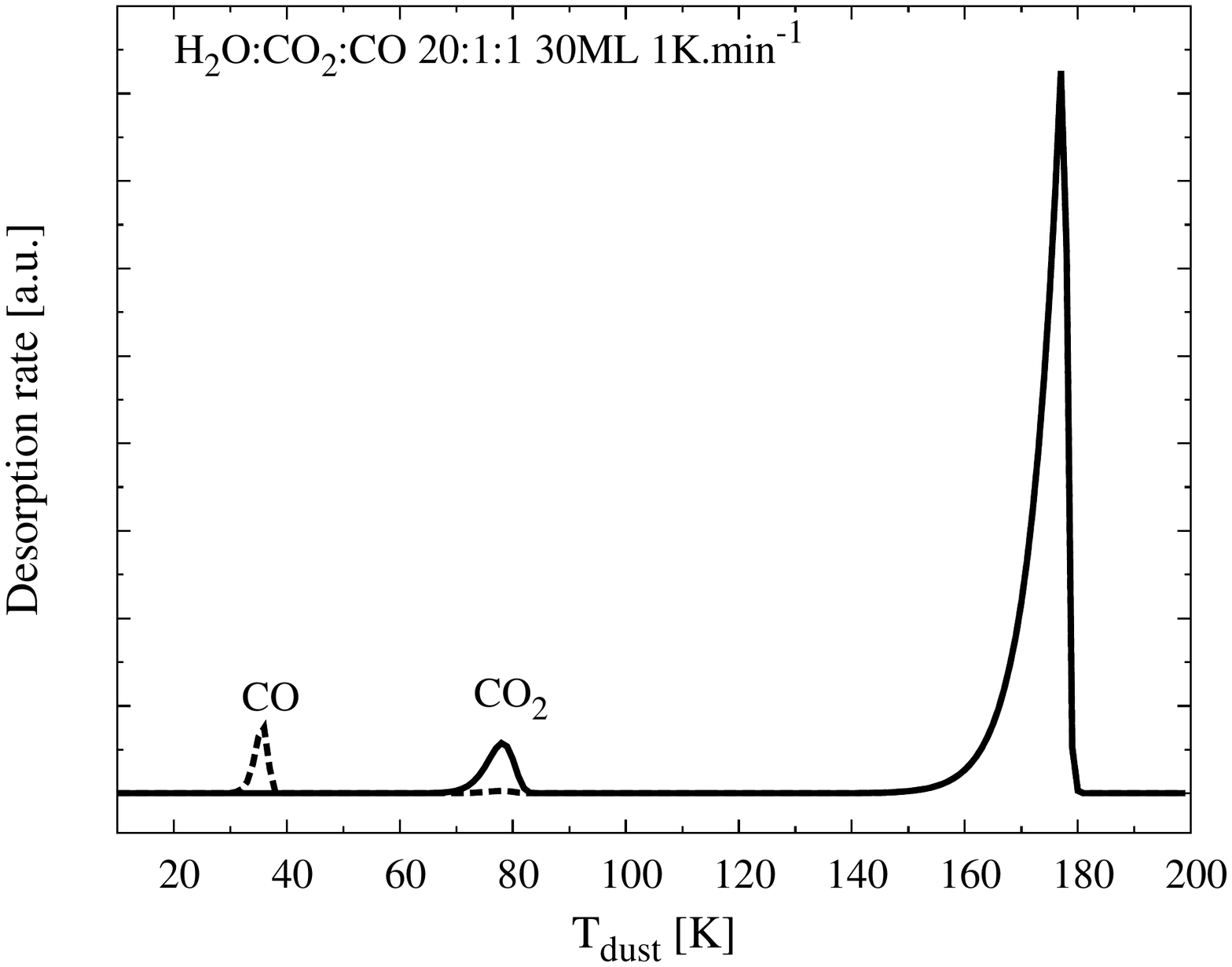}
\caption{Simulated CO and CO$_2$ desorption rates as a function of T$_\text{dust}$ from a tertiary water dominated ice mixture H$_2$O:CO$_2$:CO 20:1:1. The heating rate used is 1K.min$^{-1}$.}
\label{Fig:TPD_h2o-co2-co_20-1-1_30ML}
\end{figure}

In the following, we used experimental results obtained by \citet{Fayolle11}, in which TPD experiments were performed on CO:H$_2$O, CO$_2$:H$_2$O and CO:CO$_2$:H$_2$O ices using the CRYOPAD experimental setup. For this comparison, we focus on the results obtained in experiments No. 6, 7 and 19 of \citet{Fayolle11}. Fig. \ref{Fig:TPD_h2o-co2_30ML} and Fig. \ref{Fig:TPD_h2o-co2-co_20-1-1_30ML} present these simulated TPD experiment using the developed 3-phase model. For these simulations, we adopted the same initial conditions as those used in \citet{Fayolle11}. The temperature was linearly increased over the time, using a 1K.min$^{-1}$ heating rate. We also turned off all the chemical reactions and considered only surface and mantle diffusion as well as desorption from the surface.

As it can be seen on these two figures, the 3-phase model gives qualitatively very similar results than those obtained by the experiment of \citet{Fayolle11}. The shapes and the positions of the simulated desorption peaks show a very good agreement with the observed ones. The fact that the H$_2$O desorption peak is shifted at higher temperature in our case, comes from the higher binding energy, i.e. $E_\text{des}(\text{H}_2\text{O})=5700$K versus $E_\text{des}(\text{H}_2\text{O})=4400$K in \citet{Fayolle11}. From these experiments, the authors measured the entrapment efficiency of CO and CO$_2$ within H$_2$O ice\footnote{The authors defined the entrapment efficiency as the fraction of volatiles with respect to its initial content ice mixture, that remains trapped within the H$_2$O ice, after the mixture is heated above volatiles sublimation temperature, but below the H$_2$O sublimation temperature.}. From our simulations and following \citet{Fayolle11}, we deduce an entrapment efficiency of 9\% with respect to H$_2$O for experiment No.6, 18\% for the No.7 and 5\% for the No. 19, which is in good agreement with the measured ones.

\bsp	
\label{lastpage}
\end{document}